\documentclass[times,twocolumn,final]{elsarticle}

\usepackage{medima}
\usepackage{framed,multirow}

\usepackage{amssymb}
\usepackage{latexsym}

\usepackage{url}
\usepackage{xcolor}
\usepackage{makecell}
\usepackage{tabularx}
\newcolumntype{C}{>{\arraybackslash}X} %

\usepackage{hyperref}

\definecolor{newcolor}{rgb}{.8,.349,.1}

\journal{Medical Image Analysis}

\begin{document}

\verso{Ngo \& Nguyen \textit{et~al.}}

\begin{frontmatter}

\title{A Transformer-based Neural Language Model that Synthesizes Brain Activation Maps from Free-Form Text Queries}

\author[1]{Gia H. \snm{Ngo}\corref{cor1}\fnref{fn1}}
\ead{ghn8@cornell.edu}
\cortext[cor1]{Corresponding author}
\author[1]{Minh \snm{Nguyen}\fnref{fn1}} 
\fntext[fn1]{indicates equal contribution}
\author[2]{Nancy F. \snm{Chen}\fnref{fn2}}
\author[1,3]{Mert R. \snm{Sabuncu}\fnref{fn2}}
\fntext[fn2]{indicates equal contribution}

\address[1]{School of Electrical \& Computer Engineering, Cornell University, USA}
\address[2]{Institute of Infocomm Research (I2R), A*STAR, Singapore}
\address[3]{Radiology, Weill Cornell Medicine, USA}

\begin{abstract}
Neuroimaging studies are often limited by the number of subjects and cognitive processes that can be feasibly interrogated.
However, a rapidly growing number of neuroscientific studies have collectively accumulated an extensive wealth of results.
Digesting this growing literature  and obtaining novel insights remains to be a major challenge, since existing meta-analytic tools are constrained to keyword queries.
In this paper, we present \textit{Text2Brain}, an easy to use tool for synthesizing brain activation maps from open-ended text queries. Text2Brain was built on a transformer-based neural network language model and a coordinate-based meta-analysis of neuroimaging studies. Text2Brain combines a transformer-based text encoder and a 3D image generator, and was trained on variable-length text snippets and their corresponding activation maps sampled from 13,000 published studies.
In our experiments, we demonstrate that Text2Brain can synthesize meaningful neural activation patterns from various free-form textual descriptions.
Text2Brain is available at \url{https://braininterpreter.com} as a web-based tool for efficiently searching through the vast neuroimaging literature and generating new hypotheses.
\end{abstract}


\end{frontmatter}

\section{Introduction}
\label{sec:intro}
A rapidly growing number of functional magnetic resonance imaging (fMRI) studies have given us important insights into the mental processes that underpin behavior. However, individual studies are often 
 power-restricted~\citep{carp2012secret,button2013power}, since the number of subjects and mental processes that can be interrogated in a single experiment is limited~\citep{church2010task}.
 One approach to digest the vast literature and synthesize across many studies is to perform a meta-analysis of the reported results, such as the coordinates of the most significant effects (e.g., 3D location of peak brain activation in response to a task).
These meta-analyses usually require expert curation of relevant experiments (e.g. \citep{costafreda2008predictors,minzenberg2009meta,shackman2011integration}).
A critical technical challenge here is the consolidation of synonymous terms.
Importantly, over time, different denominations might be used in different contexts or invented to refine existing ideas.
For instance, ``self-generated thought'', one of the most highly studied functional domains of the human brain~\citep{smallwood2013distinguishing}, can be referred to by varying terms, such as ``task-unrelated thought''~\citep{andrews2014default}.

The selection of reported results for meta-analysis can be automated on data scraped from the published literature~\citep{yarkoni2011large,dockes2020neuroquery,rubin2017decoding}.
Two popular examples of this direction are Neurosynth~\citep{yarkoni2011large} and more recently Neuroquery~\citep{dockes2020neuroquery}.
Neurosynth utilizes automated keyword search to retrieve relevant studies and statistical tests to find summary brain activation maps corresponding to the keywords.
Unlike Neurosynth, Neuroquery is a predictive model that synthesizes activation maps from keywords in the input query.
Despite their differences in modeling, both Neurosynth and Neuroquery only support queries consisting predefined keywords.
Furthermore, Neurosynth does not explicitly handle long queries, while Neuroquery relies on superficial lexical similarity via word co-occurences for inference of longer or rarer queries.
We propose an alternative approach named Text2Brain, which builds on recent neural language models and permits more flexible free-form text queries.
Text2Brain captures a more fine-grained and implicit semantic similarity via vector representations from the neural language model in order to retrieve more relevant studies.
Furthermore, in contrast to tools like Neuroquery, our method computes synthesized activation maps via a 3D convolutional neural network (CNN) model, which we empirically demonstrate, can capture coarse and fine details.

We compare Text2Brain's predictions with those from Neurosynth and Neuroquery, where we used article titles as free-form queries.
Furthermore, we assess model predictions on independent test datasets, including reliable task contrasts and meta-analytic activation maps of well-studied cognitive domains predicted from their descriptions.
Our analysis shows that Text2Brain generates activation maps that better match the target images than the baselines tools.
Given its flexibility in taking input queries, Text2Brain can be used as an educational aid as well as a tool for synthesizing maps based on published results or generating novel hypotheses for future research.
Compared to our conference article~\citep{ngo2021text2brain}, we have extensively expanded our results and analysis.
Specifically, we have expanded on the model validation on article titles with a different test set (section~\ref{subsec:exp_title_setup} and~\ref{subsec:title_results}), added additional evaluation on the contrast maps predicted from their descriptions (section ~\ref{subsec:contrast_results}).
New results and discussion have also been added to this paper, including a high-level conceptual comparison of models (section~\ref{subsec:high_level_comparison}), new experiments on predicting representative meta-analytic results (section~\ref{subsec:contrast_methods} and~\ref{subsec:contrast_results}), and quantitative analysis of the models' robustness to input queries (section~\ref{subsec:synonym_methods} and~\ref{subsec:synonym_results}).

\section{Datasets and Methods}
\subsection{Model overview}
\begin{figure}
    \centering
    \includegraphics[width=\linewidth]{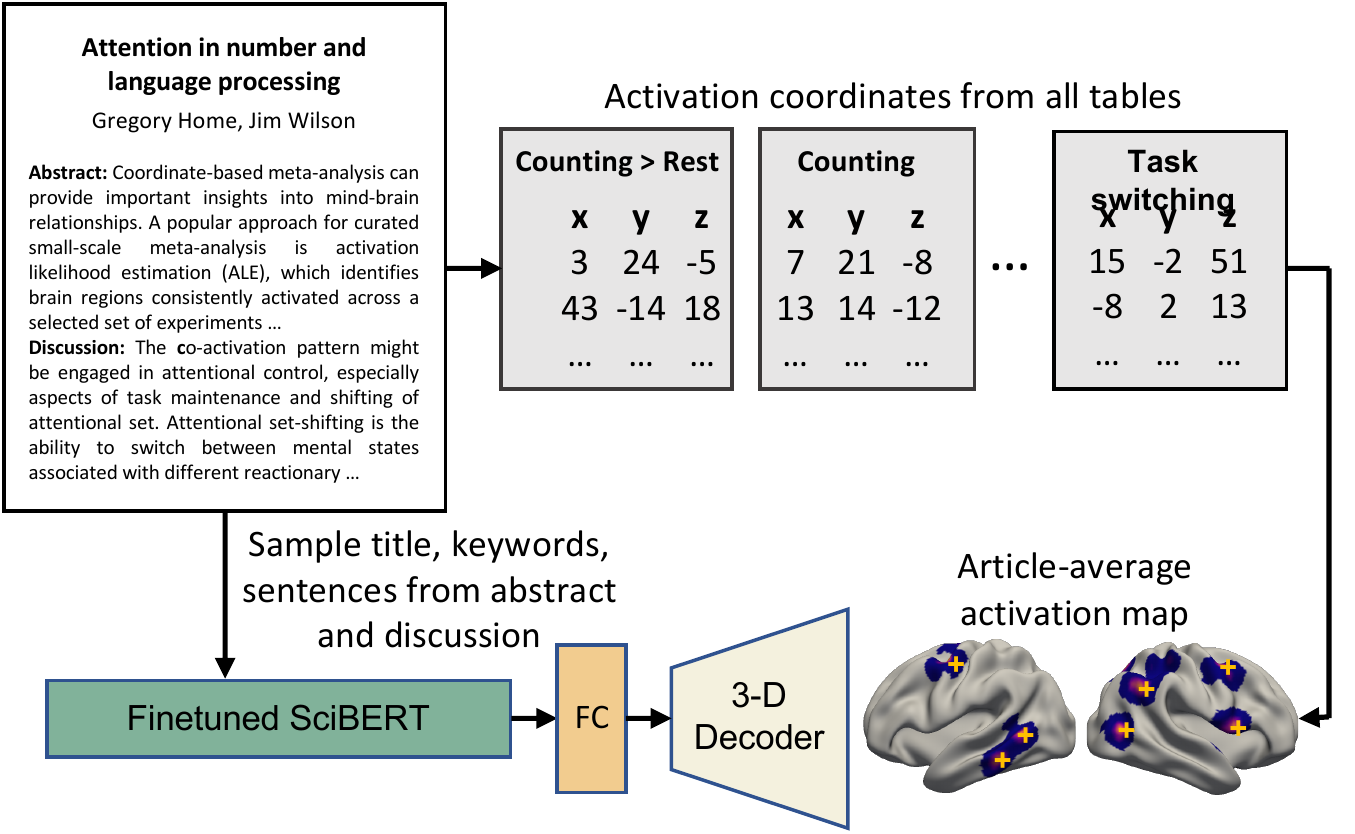}
    \caption{Overview of data preprocessing, the Text2Brain model, and training procedure. All activation maps are 3D volumes, but projected to the surface for visualization.}
    \label{fig:model}
\end{figure}

Figure~\ref{fig:model} shows the overview of this work, including data generation, model architecture, and model training.
The Text2Brain model has an encoder-decoder architecture that maps text sequences into brain activation maps~(Section~\ref{subsec:model}).
Its transformer-based encoder uses self-attention to encode a snippet of text input into vector representation~\citep{vaswani2017attention,devlin2018bert}.
Text2Brain's 3D convolutional decoder (CNN) then translates the vector representation into a 3D brain activation map.
The Transformer is currently the most effective approach for modeling text since it can capture long-distance dependency between words and can learn efficiently through self-supervision from massive text corpora~\citep{jawahar2019does,raffel2020exploring}.
On the other hand, 3D CNNs are the most dominant architectural design in medical imaging~\citep{milletari2016v,kamnitsas2017efficient}.

In our proposed approach, we first extract full text and activation coordinates from each research article to create data samples.
Each sample consists of an input snippet from the full text and an output 3D activation map created using the coordinates (Section~\ref{subsec:data}).
Text2Brain is trained to associate the input text to activations at various spatial locations.
Since Text2Brain's transformer-based encoder is context-sensitive, it can better extract information from free-form query by refining the vector representation depending on the specific phrasing of the text inputs~\citep{tenney2019you}.
In contrast, the classical keyword search mainly exploits co-occurrence of keywords regardless of context and therefore may struggle on more nuanced queries~\citep{salton1988term}.
Furthermore, keyword search approaches store one activation map for each supported keyword, which are in turn linearly combined for queries. This approach can limit how many keywords are supported~\citep{yarkoni2011large,dockes2020neuroquery}.
On the other hand, Text2Brain stores the text and activation maps content in its parameters and can scale better to diverse input queries~\citep{petroni2019language}.
We use data augmentation to encourage Text2Brain to construct and store rich many-to-one mappings between textual description and activation maps (Section \ref {subsec:training}).
This allows Text2Brain to better map semantically similar text queries to similar activation maps.

\subsection{Implementation}
\label{subsec:model}

Figure~\ref{fig:model} bottom left corner shows the Text2Brain model with its text encoder and 3D CNN image decoder.
Text2Brain's text encoder is based on SciBERT, a BERT model that has been trained using scientific articles~\citep{beltagy2019scibert}.
BERT is a transformer-based model with bidirectional self-attention trained via self-supervision to learn semantic representations of textual input~\citep{devlin2018bert}.
The text encoder outputs a vector representation of dimension $768$.
This vector is projected using a fully-connected layer and then reshaped to a 3D volume of dimension $4\times5\times4$ voxels with $64$ channels at each voxel.
The image decoder consists of 3 transposed 3D convolutional layers with 32, 16, 8 channels respectively.
Text2Brain was trained using the Adam optimizer~\citep{loshchilov2018decoupled} and the mean-squared error with a batch size of 24 for 2000 epochs.
The learning rate for the text encoder and image decoder are set at $10^{-5}$ and $3\times{10^{-2}}$ respectively.
The model's source code is available at \url{https://github.com/sabunculab/text2brain}.

\subsection{Data Preprocessing}
\label{subsec:data}
We used the same set of 13,000 neuroimaging articles previously released in~\citep{dockes2020neuroquery} in our experiments.
Each article contains one or more tables of results that reported coordinates of peak activation in MNI152 coordinate system~\citep{lancaster2007bias}.
The activation foci are also publicly released by Neuroquery~\citep{dockes2020neuroquery}.
Following the same procedure as \citep{dockes2020neuroquery}, the set of activation foci associated with each table is used to generate an activation map by placing a Gaussian sphere with full width at half maximum (FWHM) of 9mm at each of the coordinates of peak activation.
The chosen FWHM allows a fair comparison with Neuroquery\citep{dockes2020neuroquery} in our experiments, and is consistent with previous work~\citep{wager2009evaluating,yarkoni2011large,yeo2015functional}.
Supplemental section~\ref{sec:ablation_fwhm} shows an analysis of the effect of the Gaussian kernel's FWHM used for preprocessing on Text2Brain's predictive accuracy on an independent test set.
This comparison confirms that the choice of the kernel's FWHM is reasonable.
An article-average activation map is also generated by averaging the activation maps of all the tables in the article.
The text associated with the activation maps are extracted from the articles' full text.
The articles' full text are scraped using their PubMedID via the NCBI API~\footnote{\url{https://www.ncbi.nlm.nih.gov/books/NBK25501/}} and the Elsevier E-utilities API~\footnote{\url{https://dev.elsevier.com/}}.
As there may be multiple text snippets corresponding to the same activation map, the next section (Section~\ref{subsec:training}) shows how the corresponding text of an activation map is selected.

\subsection{Training}
\label{subsec:training}
Each training sample consists of a text-activation map pair and correspond to an neuroimaging article.
The activation map is sampled uniformly at random from the union set of table-specific maps and article-average map.
For each table-specific map, the first sentence of the corresponding table caption is chosen as the map's associated text.
Our initial data exploration suggested that the first sentence to be the most relevant description of the activation map.
For each article-average map, the associated text that describes the activation map is sampled uniformly at random from the following four sources: (1) the article's title; (2) one of the article's keywords; (3) the article's abstract; and (4) a randomly chosen subset of sentences from the discussion section of the article.
This data augmentation strategy encourages Text2Brain to generalize over input texts of different lengths.
Furthermore, matching the same activation pattern with multiple different text snippets encourages the model to recognize important words common across the snippets and to learn the association between different but synonymous words.
Supplemental Figure~\ref{fig:supp_ablation} shows our ablation study on the sampling strategy.
The liberal (and likely noisy) construction of image-text pairs appears to perform better than more deliberate coupling of image-text snippets strategies (not reported) that we tried in our preliminary experiments. We surmise that simply presenting different text snippets to a target brain image is analogous to another augmentation strategy that allows the neural network to pool across samples and learn the relevant words and their weights with respect to the target brain maps.
Training with the set up in~\ref{subsec:model} takes approximately 75 hours on one NvidiaRTX GPU while one inference pass with an input query of up to 140 characters takes less than 1 second.

\subsection{Baselines}
We compare Text2Brain to 2 different baselines: Neurosynth~\citep{yarkoni2011large} and Neuroquery~\citep{dockes2020neuroquery}.
For a keyword, Neurosynth first finds all neuroimaging articles that mention that keyword.
Then, one statistical test per voxel is performed across the activation maps corresponding to those studies to determine a significant association.
Since Neurosynth was not formulated to handle multiple-word queries, for  such query, we performed statistical test using activation maps from all articles that contain at least one of the keywords in the query.

Neuroquery extends Neurosynth's vocabulary of keywords by including more curated keywords from lexicons such as MeSH, NeuroNames, and NIF~\citep{lipscomb2000medical,bowden1995neuronames,gardner2008neuroscience}.
The keyword encoding is obtained after performing non-negative matrix factorization of the articles' full text (as a bag of keywords) represented with term frequency - inverse document frequency (TF-IDF) features~\citep{salton1988term}.
A ridge regression model was trained to map the text encoding to the activation.
The inference of a keyword is smoothed by a weighed average of its most related keywords (in the TF-IDF space).
For multiple-word queries, the predicted activation map is obtained by averaging the activation maps from all keywords in the input, weighed by the coefficients learnt during training.

\subsection{Evaluation Metrics}
\label{subsec:metrics}
For thresholded target activation maps such as those computed by ALE~\citep{eickhoff2009coordinate}, the predicted brain maps are thresholded to retain the same number of most activated voxels as the target.
For example, given an estimated activation map by ALE with statistically significant clusters of activation that cover 25\% of the the brain volume, the brain maps predicted by Text2Brain, Neuroquery, and Neurosynth are also thresholded to retain the top 25\% most activated voxels in each map.
The accuracy of prediction is measured by Dice score~\citep{dice1945measures} which quantifies the extent of overlap between the predicted and target brain maps (details are in Supplemental Section~\ref{subsec:supp_metrics}).

Furthermore, we use Dice scores at different thresholds to estimate the similarity between predicted and target activation maps at different levels of detail~\citep{ngo2022predicting}.
This evaluation procedure is similar to that used in \citep{dockes2020neuroquery} for a thresholded target map, but we apply the same thresholding to both the target and predicted map.
For example, at 5\% threshold (considering the 5\% most activated voxels), the Dice score measures the correspondence of the fine-grained details between the target and predicted activation maps.
At higher thresholds (e.g. 25\%), the score captures the gross agreement between activation clusters.
We also estimated  the area under the Dice curve (AUC) as a summary measure using approximated integration of Dice scores across all thresholds from 5\% up to 30\%.
Supplemental Figure \ref{fig:supp_dice} shows the Dice curve for an example pair of target-predicted activation maps.
Note that the range of thresholds in the x-axis also conveys the maximum percentage of the gray matter mask that has an activation in the target brain map.
For example, if only a proportion of gray matter mask has activation, such as the case of Neuroquery prediction that mostly extends up to 30\% of the gray matter mask or a sparse target activation pattern from the coordinate-based meta-analysis, the x-axis range will not be extended up to 1.

In our experiments, all evaluation is performed in the MNI152 volumetric space, which is the original space of all predicted maps.
For visualization, with activation maps that mostly concentrate in the cerebral cortex, the original volumetric images are transformed from MNI152 space to fs\_LR surface space using Connetome Workbench~\citep{van2013wu} via the FreeSurfer surface space~\citep{buckner2011organization,fischl2012freesurfer}, with isolated surface clusters of less than 20 vertices being removed~\citep{wu2018accurate}.
Activation maps with significant activation in the non-cortical parts of the brain are visualized by cross-sectional slices with significant activation using Nilearn~\citep{abraham2014machine}.

\subsection{High-level model comparison}
\label{subsec:high_level_comparison}
\begin{table}[]
    \centering
    \begin{tabular}{|l|l|l|l|}
    \hline
                                   & Neurosynth & Neuroquery & Text2Brain \\ \hline
         Vocabulary                & Fixed      & Fixed      & Unlimited \\ \hline
         \makecell[l]{Handle of\\complex query} & None       & \makecell[l]{Lexical\\similarity} & \makecell[l]{Semantic\\similarity} \\ \hline
         \makecell[l]{Predictive\\models}                 & None & \makecell[l]{TF-IDF,\\linear\\regression} & \makecell[l]{Transformer,\\ convolution} \\ \hline
    \end{tabular}
    \caption{High-level comparison of approaches to meta-analytic brain maps generation}
    \label{tab:comparison}
\end{table}

Text2Brain can better handle input text than prior approaches because its vocabulary is not limited to a fixed pre-defined set of words.
In contrast, Neurosynth and Neuroquery rely on fixed word vocabularies and cannot predict for queries consisting of out-of-vocabulary words.
Besides, Neurosynth's and Neuroquery's vocabularies are not sufficiently extensive, covering only a fraction (under 10\%)~\citep{dockes2020neuroquery} of terms in relevant neuroimaging lexicons such as Cognitive Atlas~\citep{poldrack2016brain} and NeuroNames~\citep{bowden1995neuronames}.
Text2Brain's usage of byte-pair encoding enables the model to handle infrequent and out-of-vocabulary words more gracefully, by breaking down those words into digestable sub-word tokens~\citep{sennrich2016neural}.
Hence, Text2Brain's vocabulary is open ended and can scale with training data to be unlimited in theory.
Besides, Text2Brain's training is not limited to only training set data.
Text2Brain can leverage self-supervised learning from non-neuroimaging scientific articles, as well as neuroimaging articles that do not report activation coordinates to learn a better text-to-activation-map transformation.
By finetuning a SciBERT text encoder pretrained on the larger dataset of scientific articles (including non-neuroimaging articles), Text2Brain seems to converge on an optimum with a more useful representational space of the input text. Supplemental section~\ref{sec:ablation_pretraining} shows the comparison between the Text2Brain model that uses pretrained SciBERT text encoder versus a randomly initialized text encoder. Evaluation on predicting article-average activation maps from both sets of test articles in the Neuroquery dataset (similar to section~\ref{subsec:exp_title_setup}) suggests that pretraining benefits the Text2Brain’s performance.
Furthermore, Text2Brain uses contextualized text embeddings to model semantic relationship between words so it can deal with nuanced queries more effectively.
Methods such as Neurosynth and Neuroquery may have difficulty dealing with complex expressions.
By simply averaging the keywords' activation maps to arrive at the prediction for a complex query, these methods may fail to account for relationship between words in the query, such as order and semantic.
Lastly, while the predictive approach of Neuroquery constructs the predicted activation map by modelling voxels' activation independently, Text2Brain generates the whole-brain activation with a 3D convolutional decoder that takes in the text encoding produced by the language model.
By upsampling and computing the whole-brain activation from a bottleneck, Text2Brain can better model both the short and long-distance relationship between voxels.

\section{Experimental Setup}
\subsection{Predict activation maps from article title}
\label{subsec:exp_title_setup}

Two test sets were created from the Neuroquery dataset of 13,000 studies.
The first test set consists of 1000 randomly sampled articles.
The second test set also consists of 1000 articles but was randomly sampled such that  the keywords (defined by the articles' authors) do not appear in the training and validation articles.
The two test sets are labeled as easy and hard test sets respectively.
Of the remaining articles, 1000 are randomly held out as a validation set for parameters tuning.
For each article, the article-average activation map is predicted from its title using Text2Brain, as well as the Neurosynth and Neuroquery baselines.
Both Text2Brain and Neuroquery were trained on the 10,000 articles in the training set.
The Text2Brain model is trained using both the articles' titles and samples from the full-text, while Neuroquery is trained on the articles' full-text.
We use predictions from the publicly available Neurosynth model at \href{https://neurosynth.org/}{https://neurosynth.org}, which was trained on the articles' abstracts.
Note that Neurosynth is not a predictive model meant for out-of-sample prediction, but for performing automated statistical testing of associations between terms and brain locations.

\subsection{Predict activation maps from contrast descriptions}
\subsubsection{Individual Brain Charting (IBC) task contrasts}
The Individual Brain Charting (IBC) project~\citep{pinho2020individual}  estimates an extensive functional atlas of the human brain via fMRI data of subjects measured under a large number of task conditions.
In particular, the IBC dataset consists of 180 task contrasts measured on 12 subjects.
We use the activation maps provided by the IBC project to measure the predictive accuracy of Text2Brain and the two baselines over a wide range of functional domains, given the contrast descriptions from IBC.

\subsubsection{Human Connectome Project (HCP) task contrasts}
While the IBC dataset offers a large number of reference brain maps, the small number of subjects might make some results less reliable.
We also utilized the Human Connectome Project (HCP) data both for reference and a measure of reliability of target maps.
The HCP dataset consists of neuroimaging data from over 1200 subjects, including task fMRI (tfMRI) of 86 task contrasts from 7 domains~\citep{barch2013function}, which overlap with 43 contrasts under the IBC dataset.
We evaluate the model prediction of HCP task contrasts from their descriptions.
While HCP provides detailed descriptions of task contrasts, we opt for the more concise contrast descriptions provided by the Individual Brain Charting (IBC) as they are more succinct and thus more favorable to the baselines.
The IBC contrast descriptions are extracted from the metadata of the activation maps released on Neurovault \href{https://neurovault.org/images/360528}{https://neurovault.org/images/360528}.
The list of all IBC description of HCP contrasts are included in Supplemental Table~\ref{tab:ibc_description}.
On the other hand, the target (ground-truth) activation maps are the HCP group-average contrast maps, as the large number of subjects provides more reliable estimates of the contrast maps.
In the analyses of this experiment, we use the agreement between the IBC and HCP maps as a measure of reliability.
Despite using similar protocols, there are subtle differences between the IBC and HCP experiments.
For instance, the original HCP language task was conducted in English but the corresponding language task in the IBC project was conducted in French.

\subsection{Predict representative meta-analytic brain maps}
\label{subsec:contrast_methods}
The automated approach to brain map generation of Text2Brain and the 2 baselines are compared against published brain maps created from a manually curated set of meta-analyses.
In particular, 5 cognitive concepts and their corresponding activation maps of 5 representative meta-analytic studies from ANIMA database~\citep{reid2016anima} were selected.
The 5 meta-analytic studies were selected for having the most number of experiments and their different coverage of the human brain.
The cognitive processes of interest are visual processing, auditory processing, motor execution~\citep{heckner2021delineating}, working memory~\citep{rottschy2012modelling}, and pain~\citep{xu2020convergent}.
Each study searches for published neuroimaging studies that contain a set of texts  queries relevant to the cognitive concept of interest.
For example, in~\cite{rottschy2012modelling}, the phrases to search for working memory-related studies are ``working memory'' and ``short-term memory''.
The same text queries for discovering relevant studies in the original meta-analysis were used as input to Neurosynth, Neuroquery, and Text2Brain.
Table~\ref{tab:domain_exp} shows the search queries and the number of experiments included in the original meta-analysis of the 5 chosen cognitive concepts.
Activation maps generated from all text input queries corresponding to each cognitive concept are averaged to yield a single brain map for each model.
The reference brain images for comparison are the activation maps released by the studies and made publicly available on ANIMA.
The reference activation maps are produced by Activation Likelihood Estimation (ALE)~\citep{turkeltaub2002meta,laird2005ale,eickhoff2009coordinate} and thresholded to retain only the statistically significant clusters of activation.
For all reference ALE maps, the cluster-level forming threshold at voxel-level is $p<0.001$ and cluster-level corrected threshold is set at $p<0.05$ by the original authors~\citep{eickhoff2012activation}.
For comparison, the generated brain maps are thresholded to keep the same number of survived voxels as those in the reference activation maps.
The accuracy of each model's generated brain map is evaluated as the Dice score between the (thresholded) generated brain map and the target (thresholded) brain map (see Section~\ref{subsec:metrics}).

\begin{table}[h!]
\centering
\begin{tabular}{ | l | l | l | } 
\hline
 Functional domain & \#Exp & Search queries \\ \hline
 \makecell[l]{Visual processing\\
 (Heckner 2021)} & 114 & \makecell[l]{visual processing\\
    face monitor\\
    face discrimination\\
    film viewing\\
    fixation\\
    flashing checkerboard\\
    passive viewing\\
    visual object identification\\
    visual pursuit\\
    visual tracking\\
    visuospatial attention} \\\hline
  \makecell[l]{Auditory processing\\
 (Heckner 2021)} & 122 & \makecell[l]{auditory processing\\
    divided auditory attention\\
    music comprehension\\
    oddball discrimination\\
    passive listening\\
    phonological discrimination\\
    pitch monitor\\
    pitch discrimination\\
    tone monitor\\
    tone discrimination} \\\hline
  \makecell[l]{Motor execution\\
     (Heckner 2021)} & 251 & \makecell[l]{motor execution\\
    writing\\
    chewing\\
    swallowing\\
    drawing\\
    isometric force\\
    motor learning\\
    grasping\\
    finger tapping\\
    button press\\
    flexion\\
    extension\\ } \\ \hline
  \makecell[l]{Working memory\\
     (Rottschy 2012)} & 189 & \makecell[l]{working memory\\
    short-term memory\\ }
    \\\hline
  \makecell[l]{Pain\\
     (Xu 2020)} & 222 & \makecell[l]{
     pain\\
    noxious\\
    nociception }
    \\\hline
\end{tabular}
\label{tab:domain_exp}
\caption{Meta-analytic studies of representative functional domains. The studies were selected from the ANIMA dataset~\citep{reid2016anima} that have the most number of experiments and covere a diverse set of brain regions.}
\end{table}

\subsection{Evaluate robustness of model prediction to semantically-equivalent queries}
\label{subsec:synonym_methods}
With the continual improvement of our understanding of the human brain and mind, neuroscientific knowledge is also an ever evolving repertoire. 
Several neuroimaging concepts have also been changing, adapting and broadening over time.
Thus, we were interested in examining if our approach is robust to semantically equivalent queries.
For example, ``self-generated thought'', one of the most intensively examined cognitive domains in neuroscience, has had its definition refined and assigned different denominations over the years. 
As a cognitive paradigm, different names have been used to refer to the set of inward-oriented psychological processes, such as ``self-generated thought''~\citep{smallwood2013distinguishing}, or ``task-unrelated thought''~\citep{andrews2014default}.
Both terms are associated with ``default network''~\citep{buckner2008brain}, the set of brain regions with elevated activation when subjects are not subjected to any external stimulus.

To assess models' prediction of synonymous queries, we utilized the ontology from the Cognitive Atlas~\citep{poldrack2011cognitive,bilder2009cognitive}.
The Cognitive Atlas is a collaborative knowledge base for neuroscience with content such as cognitive concepts, their description and synonyms (aliases) contributed by the project's voluntary participants~\citep{miller2010cognitive}.
At the time of our experiments, Cognitive Atlas includes 885 concepts with definition, 108 of which have at least one alias.
We considered a model to be robust with respect to a specific cognitive concept's definition if the activation map predicted from the description matches the predicted map from the concept's name.
In particular, given a model's predicted brain maps from all 885 Cognitive Atlas concept names and their description, we assess if the model's brain map predicted from a concept's definition is one of the $k$ maps (out of 886 possible maps) most similar to the model's brain map predicted from the concept's name.
In our experiments, top-1, top-5 and top-10 matching accuracy were evaluated using Dice AUC metrics.
The different values of $k$'s account for the uncertainty of the concepts' natural language text, e.g., different contributors might use different names to refer to the same concept.
Similarly, models' robustness with respect to a cognitive concept's alias is measured by the accuracy of matching the activation maps predicted from the text of a concept's alias and its name.

\section{Results}
\label{sec:results}
\subsection{Validation of activation maps predicted from article title}
\label{subsec:title_results}
\begin{figure}
    \centering
    \includegraphics[width=0.9\linewidth]{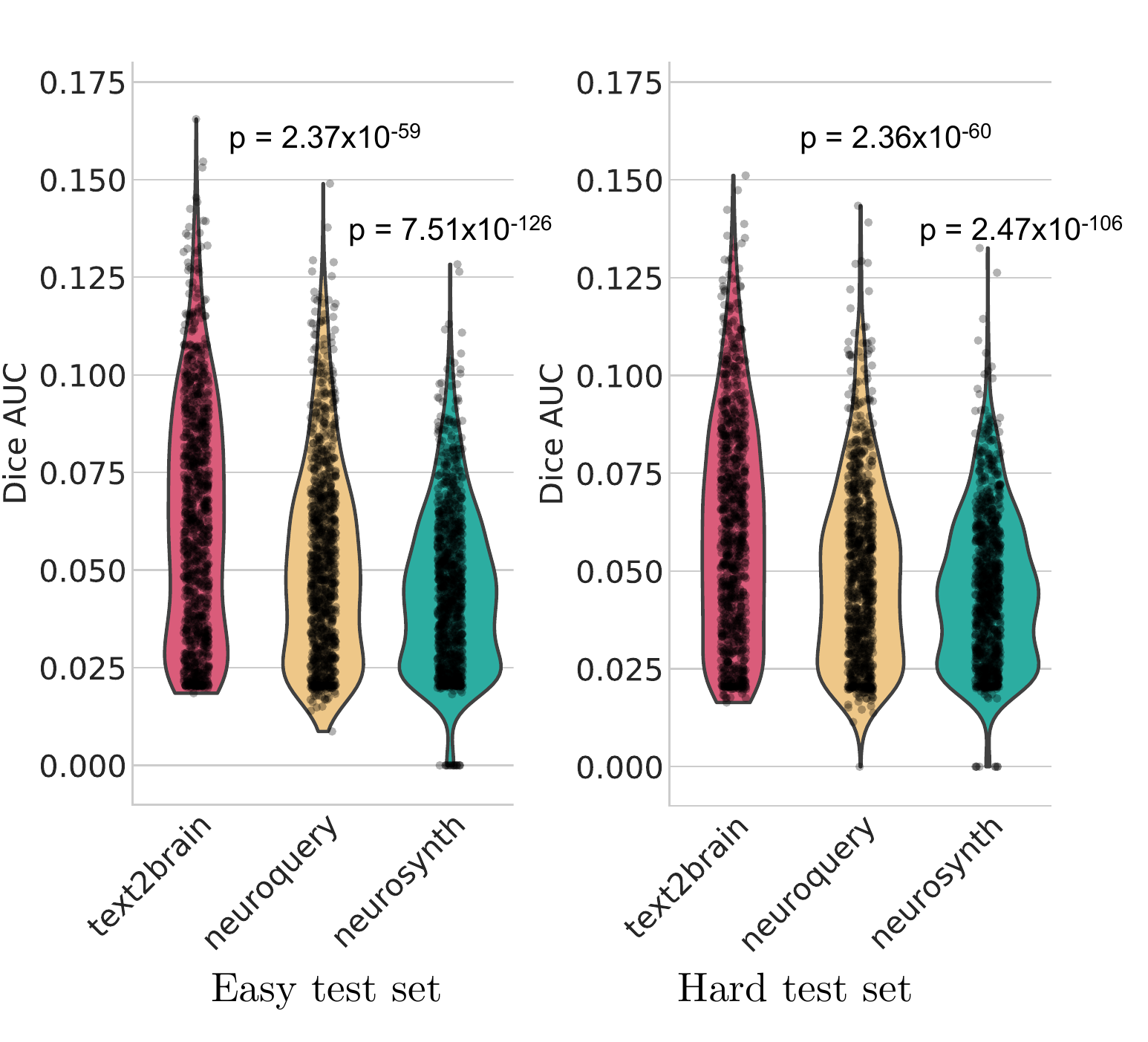}
    \caption{Evaluation of article-average activation maps predicted from their titles measured in area under the Dice curve (AUC) score. The left and right graph show the Dice AUCs of samples from the easy and hard test sets, respectively (Section~\ref{subsec:exp_title_setup}). The p-values are computed from paired-sample t-tests between Text2Brains and each of the 2 baselines.}
    \label{fig:article_AUC}
\end{figure}
Figure \ref{fig:article_AUC} shows the quality of activation maps predicted from the titles of 1000 articles in each of the two test sets (section~\ref{subsec:exp_title_setup}).
In the easy test set (the test articles' keywords can overlap with the training articles'), the proposed Text2Brain model (mean Dice AUC = $0.0636$) outperforms Neuroquery (mean Dice AUC = $0.0523$) and Neurosynth (mean Dice AUC = $0.0453$).
In the hard test set (the test articles' keywords are not present in the training set), the Text2Brain model (mean Dice AUC = $0.0609$) also performs better than Neuroquery (mean AUC = $0.0499$) and Neurosynth (mean AUC = $0.0457$).
Paired-sample t-tests show that the performance differences in both test sets are statistically very significant.
The p-values when comparing Neuroquery and Neurosynth are $p=5.25\times10^{-27}$ and $p=2.40\times10^{-12}$.
Fig.~\ref{fig:article_AUC} also indicates how the different models handle out-of-sample input text.
Text2Brain can make a prediction for all input texts, evident with positive Dice AUCs for all samples.
On the other hand, Neurosynth fails to make prediction for some article titles in both test sets, resulting in zero Dice AUCs for such samples.
Similarly, Neuroquery fails to make prediction for some samples in the hard test set.
These failure cases are caused by the limited vocabularies of Neurosynth and Neuroquery that cannot cover the words in the test input queries.
On the other hand, the language model of Text2Brain is finetuned from SciBert, which has been pretrained on a broader lexicon and utilizes sub-word tokens to extend the vocabulary to unseen words (more details in Section~\ref{subsec:high_level_comparison}).

\subsection{Prediction of task contrast maps from description}
\label{subsec:contrast_results}
\begin{figure}
    \centering
    \includegraphics[width=\linewidth]{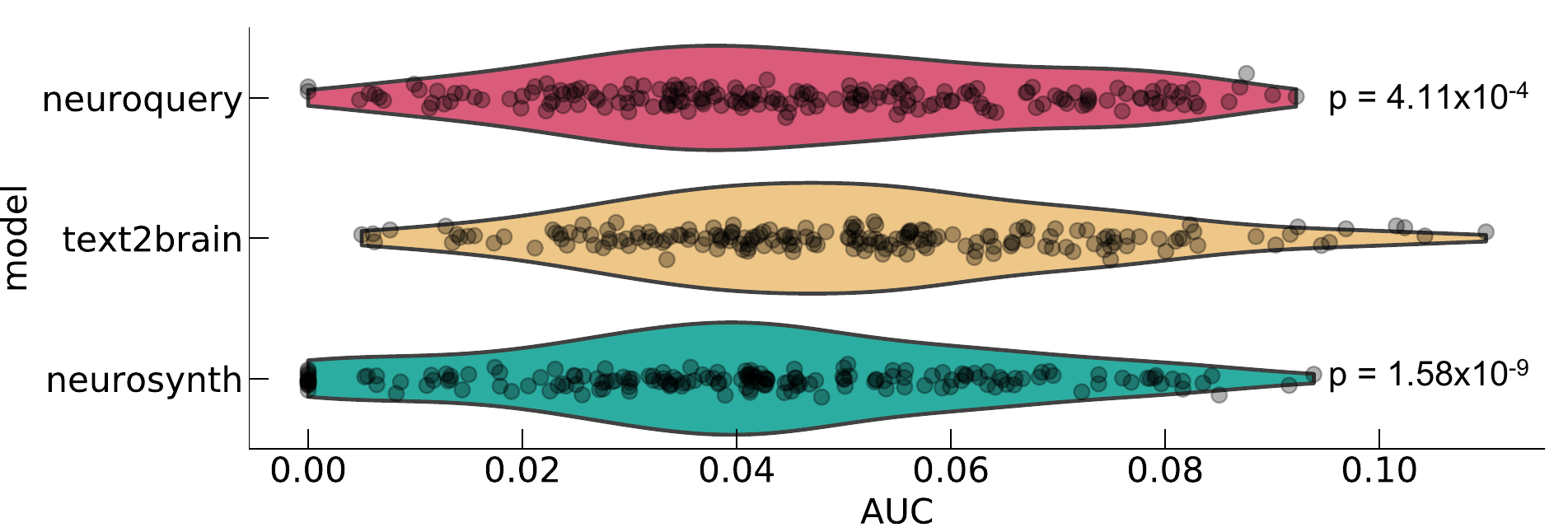}
    \caption{Dice AUCs of predicted IBC task activation maps from contrasts' description. The p-values are estimated from paired-sample t-tests between Text2Brain against the two baselines.}
    \label{fig:ibc_auc}
\end{figure}
Fig.~\ref{fig:ibc_auc} shows the Dice AUC scores for the prediction of Text2Brain, Neuroquery and Neurosynth against the IBC group-average task contrast maps.
Text2Brain (mean Dice AUC = 0.0507) improves upon both Neuroquery (mean Dice AUC = 0.0457, $p = 4.11\times{10^{-4}}$), and Neurosynth (mean Dice AUC = 0.0404, $p=1.58\times{10^{-9}}$).
The p-values are measured by 2-tail paired-sample t-test between Text2Brain and the two baselines.

\begin{figure}
    \centering
    \includegraphics[width=\linewidth]{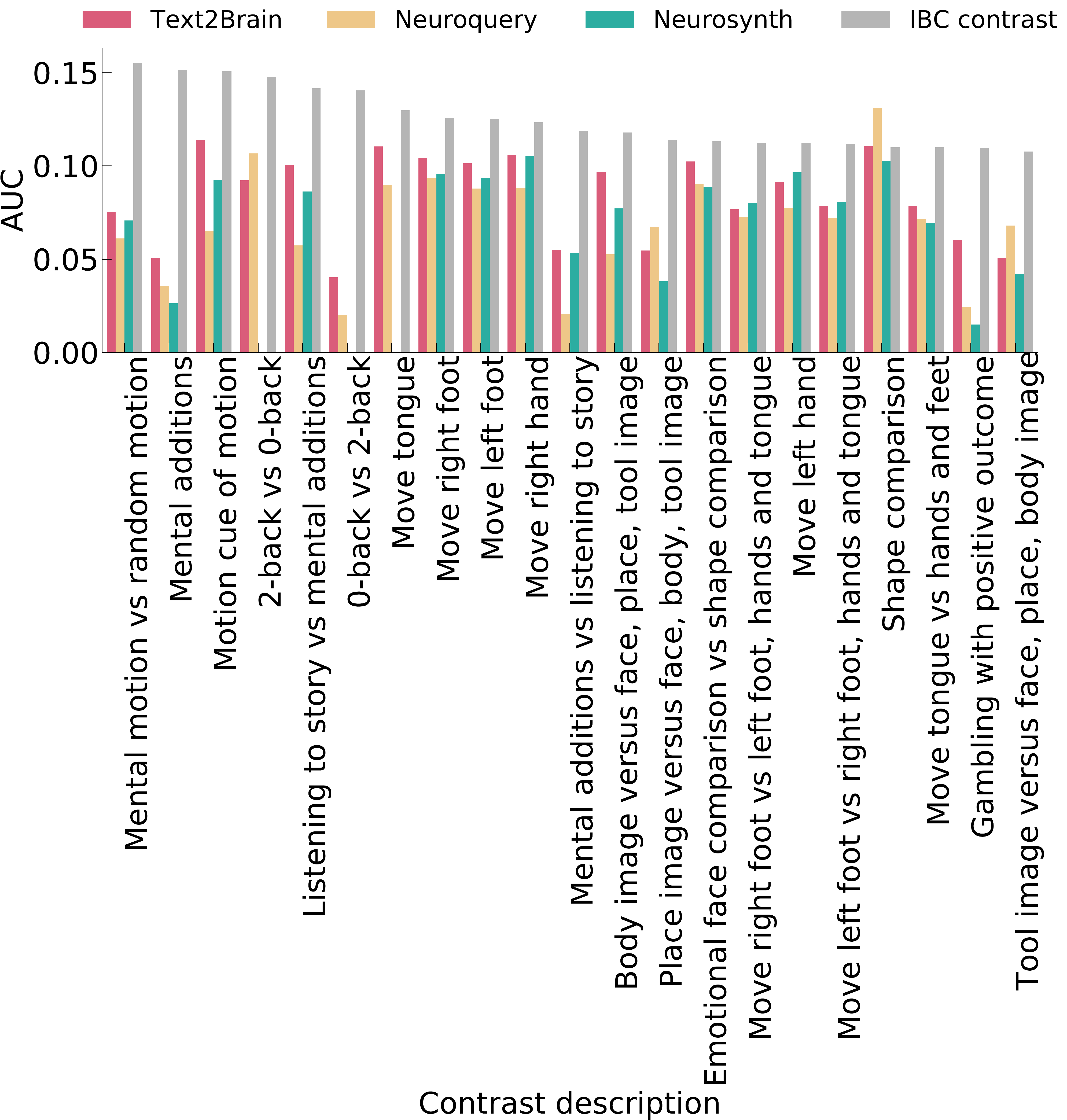}
    \caption{Dice AUCs of predicted HCP task activation maps from contrasts' description. The graph includes 22 contrasts with the highest HCP-IBC's Dice AUC scores and sorted in decreasing order.}
    \label{fig:hcp_auc}
\end{figure}
Fig.~\ref{fig:hcp_auc} shows the AUC scores for the prediction of the three models and the IBC average contrasts, against the HCP target maps.
The 22 contrasts with above-average HCP-IBC's AUC scores, considered to be the reliable contrasts, are shown.
Across all 43 HCP contrasts, Text2Brain (mean AUC = $0.082$) performs better than the baselines, i.e. Neuroquery (mean AUC = $0.0755$, $p = 0.08$), Neurosynth (mean AUC = $0.047$, $p = 1.5\times10^{-5}$), where $p$-values are computed from the paired t-test between Text2Brain's and the baselines' prediction.
As reference, IBC contrasts yield a mean AUC = $0.094$ when compared to the corresponding HCP maps (Statistical comparison with Text2Brain, $p = 0.077$).
\begin{figure*}
    \centering
    \includegraphics[width=\linewidth]{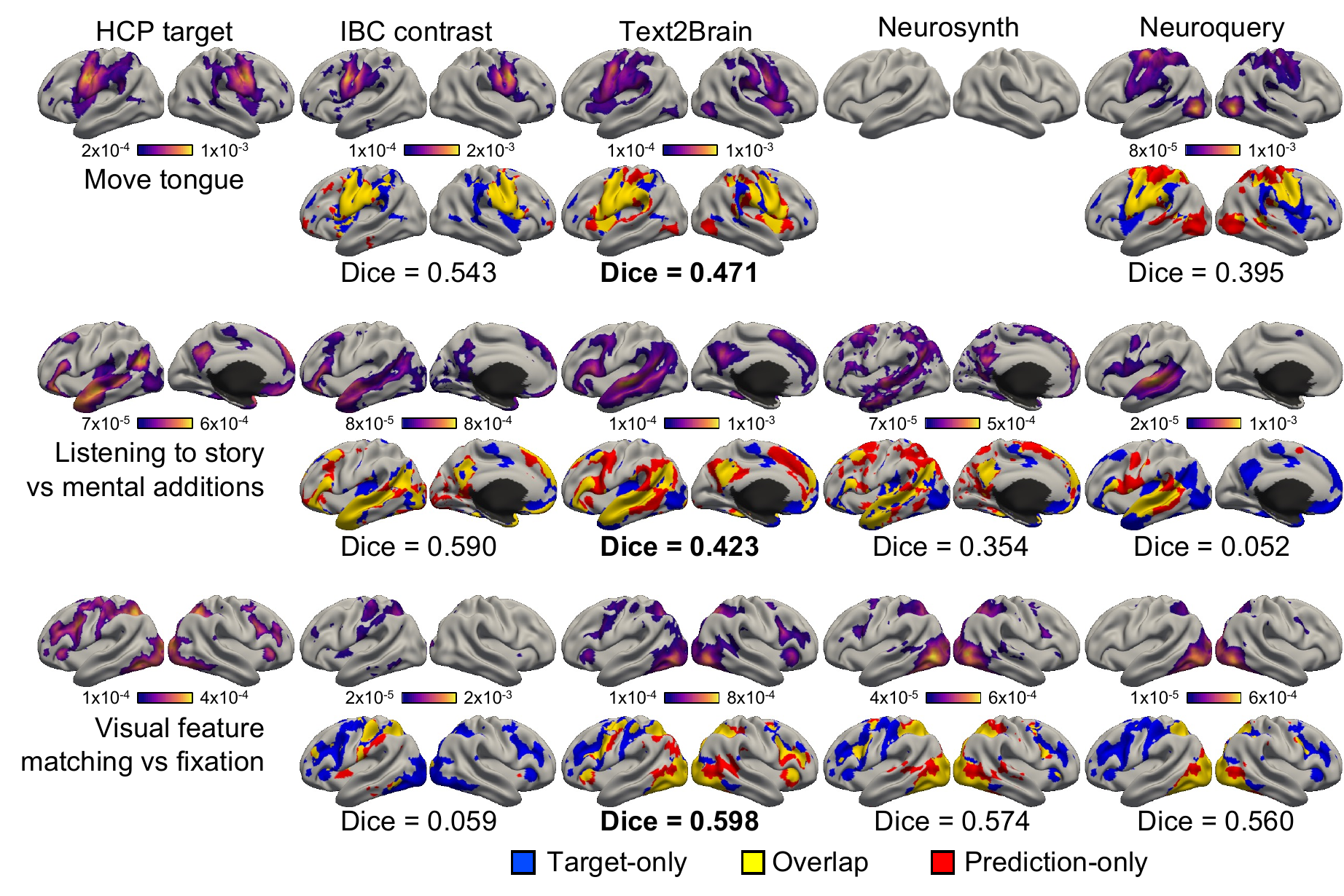}
    \caption{Task activation maps predicted from contrasts' description. Each row shows both the thresholded maps of the top 25\% most activated voxels (top) and the overlap between predicted and target binarized brain maps. Blue is activation in the target contrast, red is the predicted activation and yellow is the overlap.}
    \label{fig:hcp_example}
\end{figure*}

Figure \ref{fig:hcp_example} shows the prediction for three contrasts correspond to different HCP task groups, namely ``MOTOR'', ``LANGUAGE'', ``RELATIONAL'' thresholded at the top 25\% most activated voxels.
The three task groups were chosen to show results for a range of target images with different levels of reliability.
The two task groups ``MOTOR'' and ``LANGUAGE'' are the two most reliable task (having the highest average HCP-IBC AUC across all contrasts), while ``RELATIONAL'' has the lowest average HCP-IBC AUC.
Text2Brain's prediction improves over the baselines for the three contrasts.
Neurosynth was not able to generate activation maps for one of the contrast descriptions (``Move tongue'').
On the other hand, for the ``Move tongue'' contrast, Neuroquery predicts activation in the primary cortex, but the peak is in the wrong location, shifted more toward the hand region of the homunculus. Additionally, there is a false positive prediction in the occipital cortex, which might be an artifact from modeling brain activation coupled with visual stimuli-related words describing the motor experiments.

\subsection{Prediction of brain maps from representative meta-analytic studies}
\begin{figure*}
    \centering
    \includegraphics[width=\linewidth]{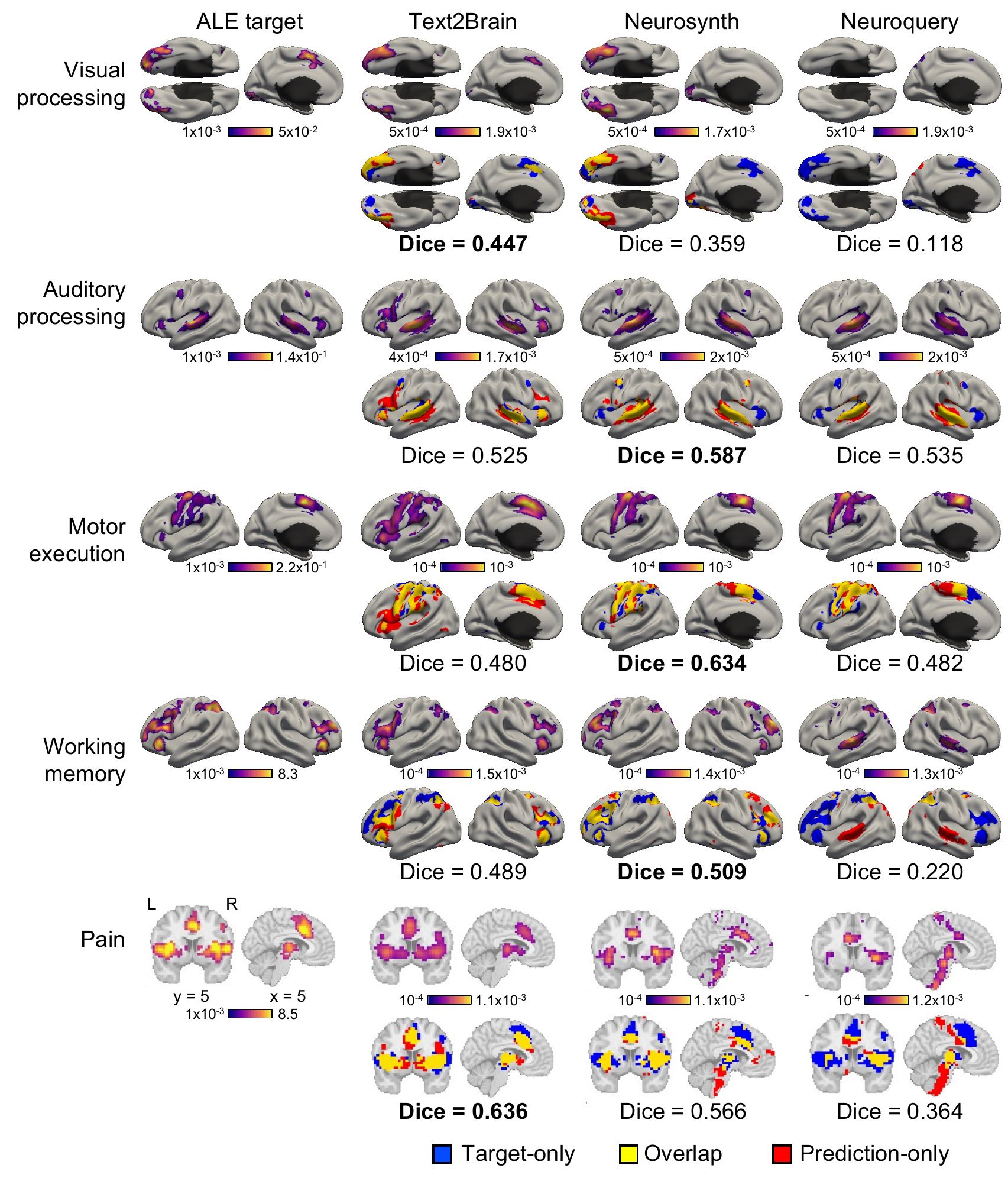}
    \caption{Prediction of brain maps from meta-analytic studies of representative functional domains. The information of the investigated functional domains are listed in Table~\ref{tab:domain_exp}. Reference and predicted activation maps of the first 4 function domains are visualized on the brain surface. The last domain (``pain'') is visualized in the volume as most activation concentrates in the non-cortical parts of the brain. For all functional domains, Text2Brain generates reasonable activation maps and comparable with the baselines for the common functional domains.}
    \label{fig:example_meta}
\end{figure*}

Figure~\ref{fig:example_meta} shows the prediction of activation maps for 5 representative meta-analytic studies with the most number of experiments from ANIMA~\citep{reid2016anima}.
Among the three models, Neuroquery has the lowest Dice score on average, with prediction on ``Visual processing'', ``Working memory'', and ``Pain'' that significantly deviates from the target maps.
On the other hand, Neurosynth-derived brain maps consistently match well against the target maps.
The high accuracy of Neurosynth prediction is expected since the five chosen cognitive concepts are among the most commonly studied concepts with the most number of experiments reporting activation coordinates in the literature.
Given high number of available experiments and the input queries mostly exist in Neurosynth's predefined keyword set, the activation coordinates scraped by automated method by Neurosynth would be very similar to the manually curated data in the original meta-analysis.
Lastly, Text2Brain also predicts consistently reasonable brain maps for all five cognitive concepts, and matches the target maps better than Neurosynth for ``Visual Processing'' and ``Pain''.
Results in Figure~\ref{fig:example_meta} shows that Text2Brain could learn appropriate relationship between common search phrases and the activation pattern of a diverse set of functional domains.

\section{Robustness of models to input queries}
\subsection{Example of ``self-generated thought'' synonyms}
We examine the prediction for ``self-generated thought'', which is one of the most extensively investigated functional domains, due to its involvement in a wide range of cognitive processes that do not require external stimuli~\citep{andrews2014default},
and is associated with the default network~\citep{buckner2008brain}.
The ground-truth map for self-generated thought, taken from~\citep{ngo2019beyond}, is estimated using activation likelihood estimation (ALE)~\citep{eickhoff2009coordinate} applied on activation foci across 167 imaging studies of 7 tasks selected based on strict criteria~\citep{spreng2009common,mar2011neural,sevinc2014contextual}.
The resulting ALE map is thresholded with the cluster-level forming threshold at voxel-level $p<0.001$, and cluster-level corrected threshold $p<0.05$~\citep{eickhoff2012activation}.%
\begin{figure*}[h!]
    \centering
    \includegraphics[width=\linewidth]{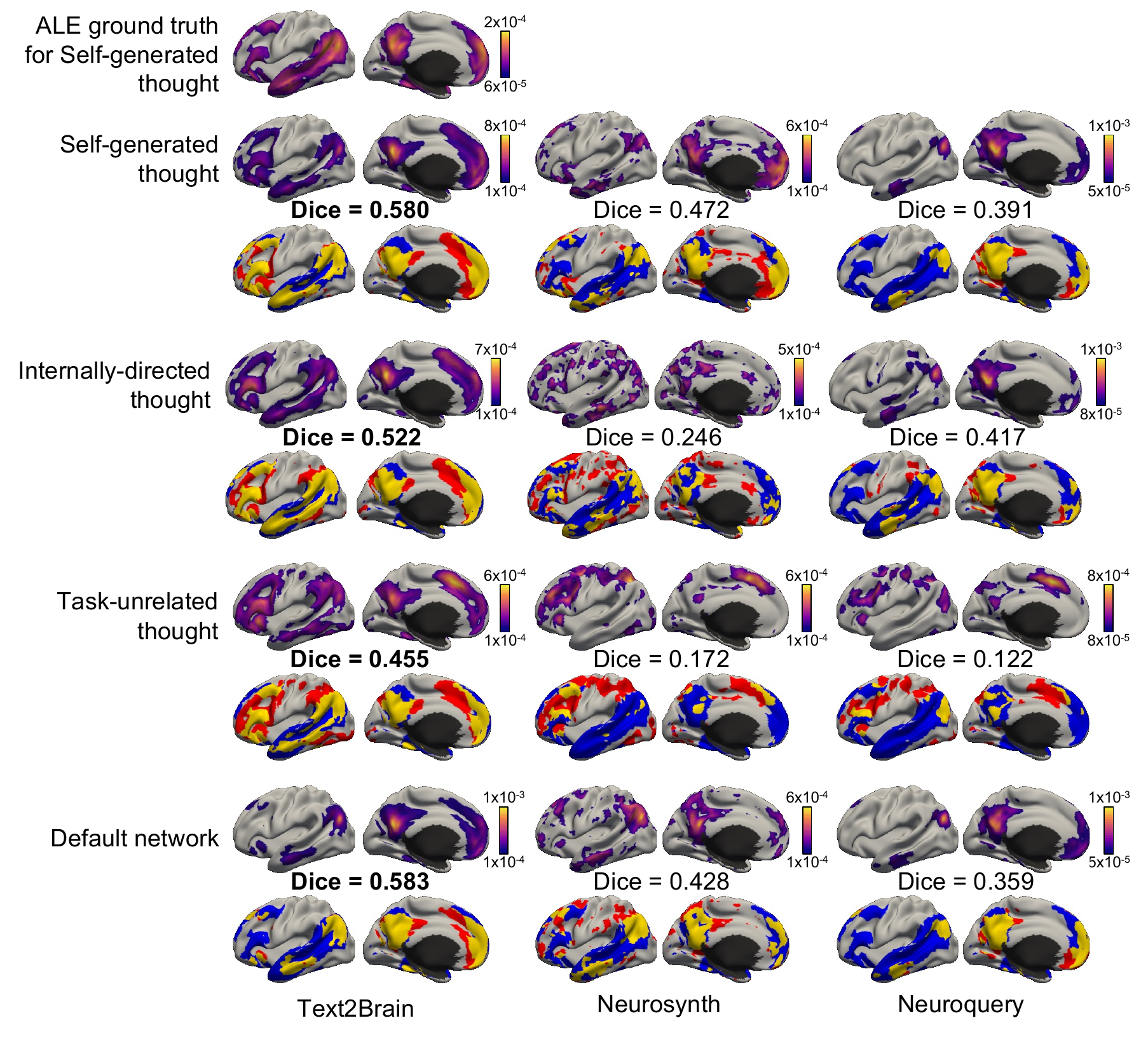}
    \caption{Prediction of self-generated thought activation map using synonymous queries. While Text2Brain generates consistent prediction across the similar queries, Neurosynth and Neuroquery's  prediction deteriorate on the ``internally-directed thought'' and ``task-unrelated thought'' queries.}
    \label{fig:self_generated_thought}
\end{figure*}
Figure \ref{fig:self_generated_thought} shows the prediction of self-generated thought activation map using four different query terms, thresholded to retain the same number of activated voxels as the target map.

Across all four queries, Text2Brain's prediction best matches the ground-truth activation map compared to the baselines.
For the ``self-generated thought'' and ``default network'' queries, all approaches generate activation maps that are consistent with the ground-truth, which includes the precuneus, the medial prefrontal cortex, the temporo-parietal junction, and the temporal pole.
Text2Brain and Neuroquery both make reasonable prediction from the ``internally-directed thought'' query while Neurosynth's prediction is largely scattered and does not match the target map.
Lastly, Text2Brain can also replicate a similar activation pattern to the target from the query ``task-unrelated thought'', evident by only a slight drop in the Dice score.
However, Neuroquery and Neurosynth both generate activation maps that differ from the typical default network's regions, such as activation in the prefrontal cortex, and also result in a large drop of the Dice scores.

\subsection{Prediction of Cognitive Atlas concepts from synonymous queries}
\label{subsec:synonym_results}
\begin{figure}
    \centering
    \includegraphics[width=\linewidth]{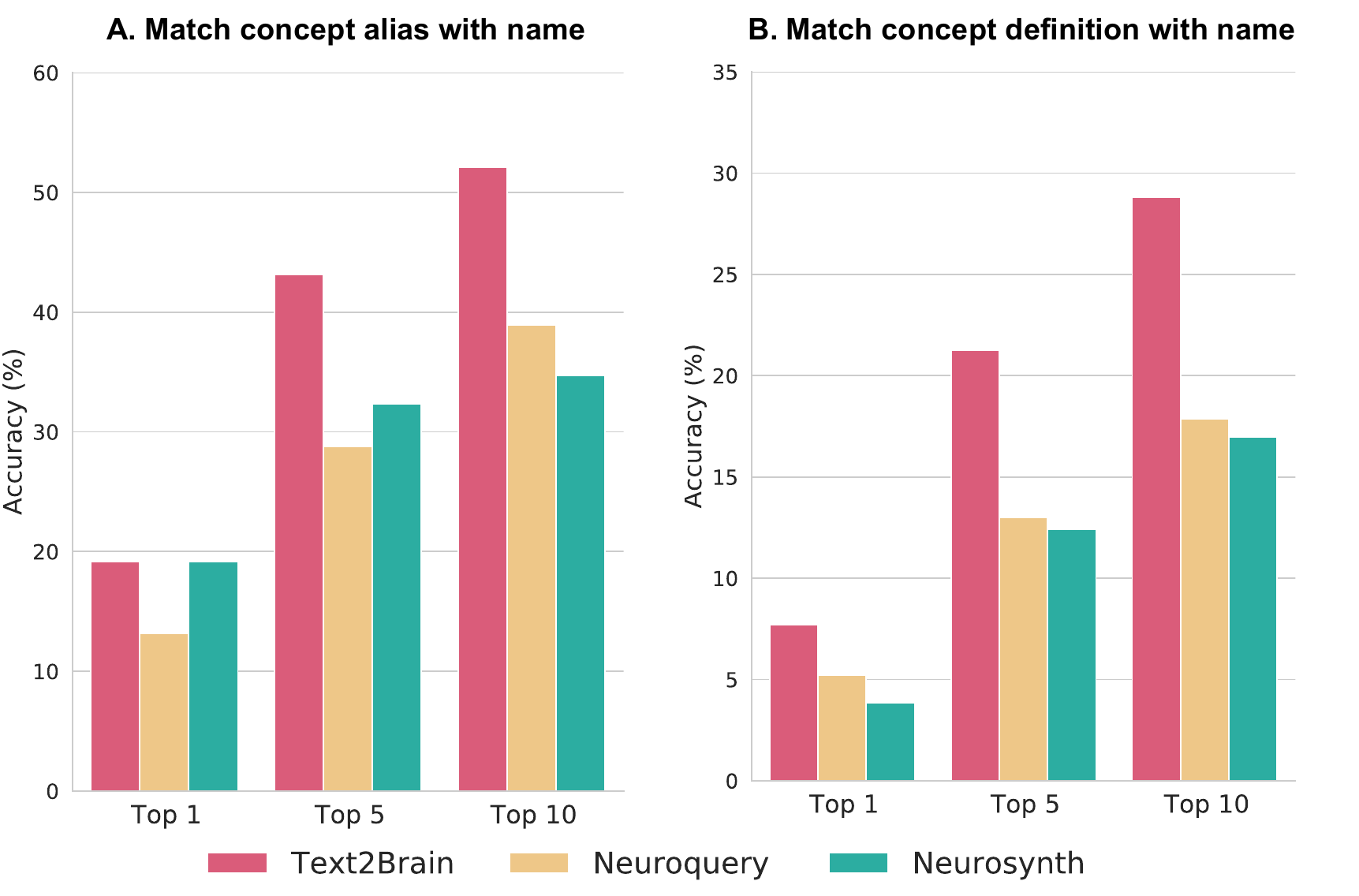}
    \caption{Accuracy of matching Cognitve Atlas concept names with their description and aliases using models' predicted brain maps.}
    \label{fig:cognitive_atlas_acc}
\end{figure}
Figure~\ref{fig:cognitive_atlas_acc} shows the accuracy of matching cognitive concept names from the Cognitive Atlas~\citep{poldrack2011cognitive} with their definitions and atlases using the different models' predicted brain maps.
Prediction by Text2Brain is more robust than both Neuroquery and Neurosynth with respect to the concept definition and alias.
In particular, Text2Brain has the same top-1 accuracy of matching the brain map predicted from a concept's alias with the prediction from the concept name compared to Neurosynth.
This result is expected given that Neurosynth can yield accurate brain map for keywords that are included in their vocabulary.
In contrast, Text2Brain improves over Neurosynth for top-1 accuracy of matching concept name with the longer text of concept definition.
Text2Brain is more robust than both Neurosynth and Neuroquery baselines in terms of top-5 and top-10 matching accuracies for both concept aliases and definitions.
Figure~\ref{fig:cognitive_atlas_acc} indicates that Text2Brain prediction is robust to natural language text queries of different length and complexity.

\section{Conclusion}
In this work, we present a model named Text2Brain for generating activation maps from free-form text query.
By finetuning a high-capacity SciBert-based text encoder to predict coordinate-based meta-analytic maps, Text2Brain captures the rich relationship in the language representational space, allowing the model to generalize its prediction for synonymous queries.
This is evident in the better performance of Text2Bran in predicting the self-generated thought activation map using different descriptions of the functional domain.
Text2Brain's capability to implicitly learn relationships between textual terms and images ensures the model can remain relevant and useful even as neuroimaging literature continues to evolve with new discoveries and rephrasing of  existing concepts.
We also show that Text2Brain accurately predicts most of the task contrasts included in the IBC and HCP dataset, validating its capability to make prediction for longer, arbitrary queries.
Text2Brain also preempts failure cases in Neurosynth and Neuroquery, where they cannot predict input queries undefined in the vocabulary list, even though these queries are relevant to neuroscience research (e.g. title of an article).
On the other hand, we also observed that Text2Brain had difficulties handling queries that involve logical reasoning, such as the direction of a contrast.
For example, while queries such as ``A vs B'' and ``B vs A'' can be inferred by human to correspond with inverted activation maps, Text2Brain sometimes treats one direction to be the same as the other.
We suspect that this type of error is likely due to the model's inability to generalize ``vs'' as an ``subtractive'' operator. Resolving such limitation will likely require modifications to the language model.
Furthore, in the future, we plan to enhance the interpretability of our approach, such as to attribute regions of activations in the generated map to specific words in the input query, as well as to efficiently match activation maps and scientific descriptions most relevant to the synthesized images.

We believe that the flexibility of Text2Brain can significantly lower the barrier for researchers at all stages of their careers to synthesize brain activation maps needed for their research.
For example, the ability of Text2Brain to generate meaningful neural activation patterns of synonymous queries for a functional domain can improve the accuracy of delineating region-of-interests (ROIs) relevant to the functional process, as well as to assess the reliability of each ROI.
Discovery of these ROIs is useful for several applications such as meta-analytic connectivity modeling (MACM)~\citep{laird2013networks}. 
We look forward to such application of Text2Brain in aiding future neuroscientific research.

\section*{Acknowledgement}
This work was supported by NIH grants R01LM012719, R01AG053949, the NSF NeuroNex grant 1707312, the NSF CAREER 1748377 grant and Jacobs Scholar Fellowship.

\bibliographystyle{model2-names.bst}\biboptions{authoryear}
\bibliography{refs}

\clearpage
\section*{Supplementary Material}
\subsection{IBC description of HCP task contrasts}
\label{subsec:ibc_description}
\begin{table*}[h]
\centering
\begin{tabular}{ | l | l | l | } 
\hline
 Task & Contrast label & Contrast description \\ \hline
 \multirow{4}{*}{LANGUAGE} & MATH & Mental additions \\
          & STORY & Listening to story \\
        & MATH-STORY & Mental additions vs listening to story \\
        & STORY-MATH & Listening to story vs mental additions \\ \hline
\multirow{3}{*}{RELATIONAL} & MATCH & Visual feature matching vs fixations \\
 & REL & Relational comparison vs fixation \\
 & REL-MATCH & Relational comparison vs matching \\ \hline
\multirow{3}{*}{SOCIAL} & RANDOM & Random motion vs fixation \\
 & TOM & Mental motion vs fixation \\
 & TOM-RANDOM & Mental motion vs random motion \\ \hline
\multirow{3}{*}{EMOTION} & FACES & Emotional face comparison \\
 & SHAPES & Shape comparison \\
 & FACES-SHAPES & Emotional face comparison vs shape comparison \\ \hline
\multirow{14}{*}{WM} & 2BK BODY & Body image 2-back task vs fixation \\
 & 2BK FACE & Face image 2-back task vs fixation \\
 & 2BK PLACE & Place image 2-back task vs fixation \\
 & 2BK TOOL & Tool image 2-back task vs fixation \\
 & 0BK BODY & Body image 0-back task vs fixation \\
 & 0BK FACE & Face image 0-back task vs fixation \\
 & 0BK PLACE & Place image 0-back task vs fixation \\
 & 0BK TOOL & Tool image 0-back task vs fixation \\
 & 0BK-2BK & 0-back vs 2-back \\
 & 2BK-0BK & 2-back vs 0-back \\
 & BODY-AVG & Body image versus face, place, tool image \\
 & FACE-AVG & Face image versus body, place, tool image \\
 & PLACE-AVG & Place image versus face, body, tool image \\
 & TOOL-AVG & Tool image versus face, place, body image \\ \hline
\multirow{11}{*}{MOTOR} & CUE & Motion cue of motion \\
 & LF & Move left foot \\
 & LH & Move left hand \\
 & RF & Move right foot \\
 & RH & Move right hand \\
 & T & Move tongue \\
 & LF-AVG & Move left foot vs right foot, hands and tongue \\
 & LH-AVG & Move left hand vs right hand, feet and tongue \\
 & RF-AVG & Move right foot vs left foot, hands and tongue \\
 & RH-AVG & Move right hand vs left hand, feet and tongue \\
 & T-AVG & Move tongue vs hands and feet, Move left hand \\ \hline
\multirow{3}{*}{GAMBLING} & PUNISH & Negative gambling outcome \\
 & REWARD & Gambling with positive outcome \\
 & PUNISH-REWARD & Negative versus positive gambling outcome \\ \hline
\end{tabular}
\label{tab:ibc_description}
\caption{Individual Brain Charting (IBC) description of Human Connectome Project (HCP) task contrasts}
\end{table*}
\clearpage
\subsection{Evaluation Metrics}
\label{subsec:supp_metrics}
Dice score~\citep{dice1945measures} is used to measure the extent of overlap between a predicted activation map and the target activation map at a given threshold.
At a given threshold of $x\%$, Dice score is computed as:
\begin{equation}
    Dice(x) = \frac{2|Prediction(x) \cap Target(x)|}{|Prediction(x)|+|Target(x)|},
    \label{eq:dice}
\end{equation}
where $|Prediction(x)|$ denotes the number of top $x\%$ most activated voxels in the predicted activation map, $|Target(x)|$ denotes the number of top $x\%$ most activated voxels in the target map, and $|Prediction(x)\cap Target(x)|$ denotes the number of voxels that overlap between the predicted and target map at the given threshold.

\begin{figure}[h]
\includegraphics[width=\linewidth]{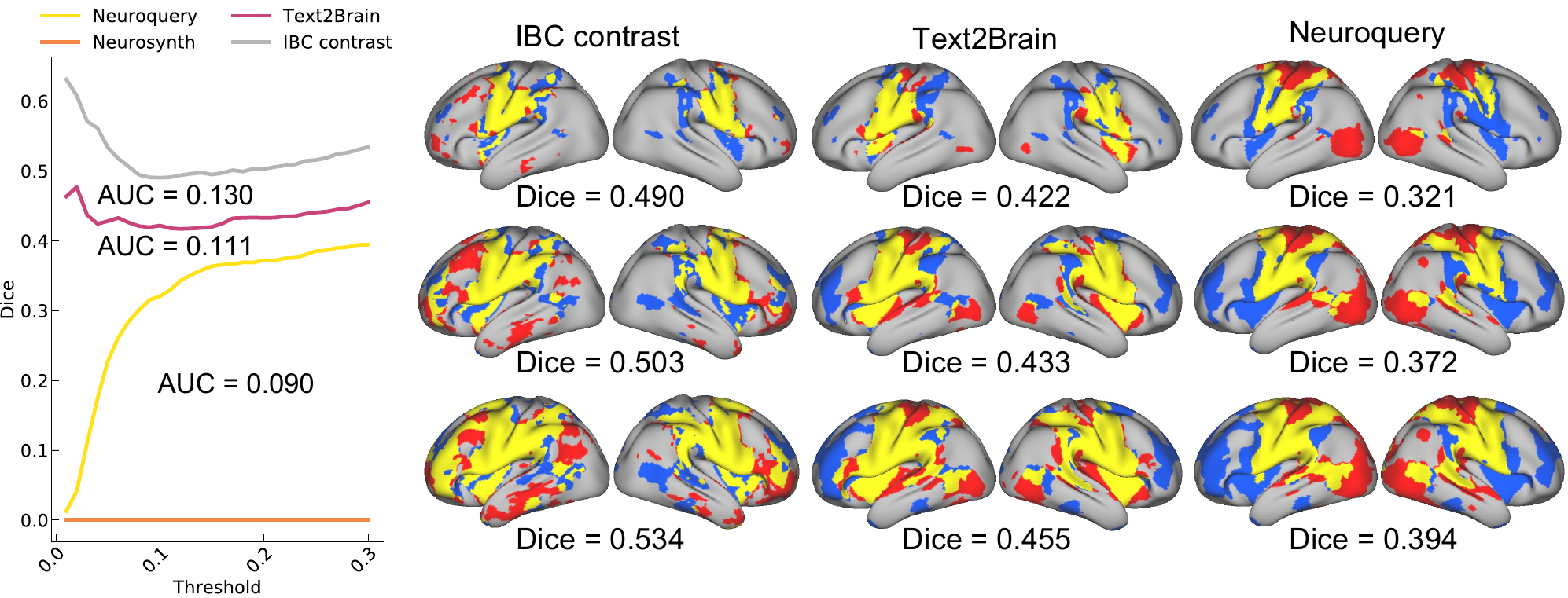}
\caption{Example Dice scores evaluated on the ``Move tongue'' contrast (in Fig.\ref{fig:hcp_example}.
The graph on the left shows the Dice scores computed between the target HCP activation map and Text2Brain's, Neurosynth's, Neuroquery's prediction, and the IBC contrast across thresholds ranging from 5\% to 30\%.
Note that Neurosynth's Dice scores are all zeros as it fails to make an inference for the input text.
The area under the Dice curve (AUC) was computed as the summary metrics of accuracy across all thresholds (e.g. Fig~\ref{fig:hcp_auc}).
The Dice AUC by Neuroquery, Text2Brain, and IBC are 0.090, 0.111, 0.130, respectively.
The brain maps on the right are visualization of the extent of overlaps between predicted and target maps at 10\%, 20\% and 30\% threshold of most activated voxels. Blue indicates activation in the target contrast, red is the predicted activation and yellow is the overlap.
}\label{fig:supp_dice}
\end{figure}

\subsection{Ablation study of sampling strategy}
\label{sec:ablation}

\begin{table}[h]
\begin{tabular}{ | l | l | } 
\hline
 Text samples                                             & Mean AUC \\ \hline
 Title + Table caption                                    & 0.0648    \\
 Title + Abstract + Table caption                         & 0.0616    \\
 Discussion + Abstract                                    & 0.0631    \\
 Discussion + Abstract + Keywords                         & 0.0651    \\
 \makecell[l]{\textbf{Title + Abstract + Keywords +}\\ \textbf{Discussion + Table caption}} & 0.0663     \\
 \hline
\end{tabular}
\caption{Performance of different sampling strategies in predicting article-average activation maps from the articles’ titles in the validation set}
\end{table}

\begin{figure}[ht!]
 \includegraphics[width=\linewidth]{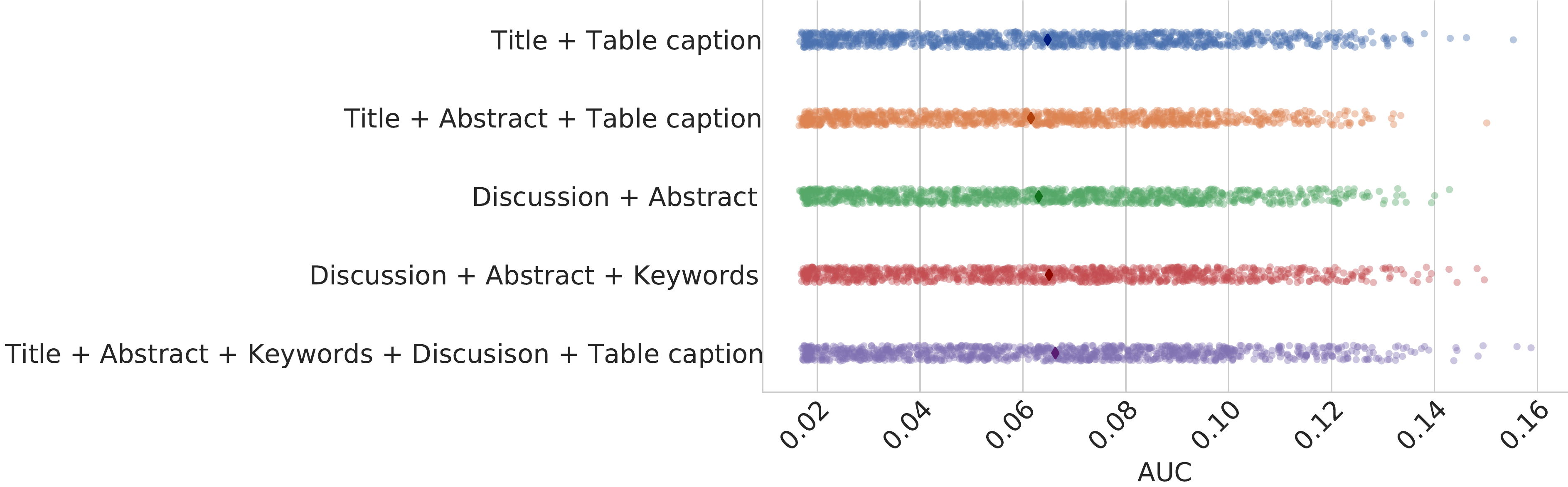}
 \caption{Performance of different sampling strategies in predicting article-average activation maps from the articles' titles in the validation set.
 All sampling strategies used the same model described in~\ref{subsec:model}.
 The model parameters used for evaluation were chosen at the epoch with the best performance on the validation set.}
 \label{fig:supp_ablation}
\end{figure}

\subsection{Ablation study of text encoding pretraining}
\label{sec:ablation_pretraining}

\begin{table}[h]
\begin{tabular}{ | l | l | l | } 
\hline
 Strategy                                                 & \makecell[l]{Easy test\\Dice AUC} & \makecell[l]{Hard test\\Dice AUC} \\ \hline
 \makecell[l]{\textbf{With text encoder} \textbf{pretraining}}          & 0.0664                               &  0.0609 \\ \hline
 \makecell[l]{No encoder pretraining}                                   & 0.0603                               & 0.0581 \\ \hline
 \makecell[l]{No encoder + tokenizer \\pretraining}                       & 0.0601                               &  0.0580 \\
 \hline
\end{tabular}
\caption{Effect of pretraining on predicting article-average activation maps from the articles’ titles in the two test sets of Section 3.1.}
\end{table}

\begin{figure}[ht!]
 \includegraphics[width=\linewidth]{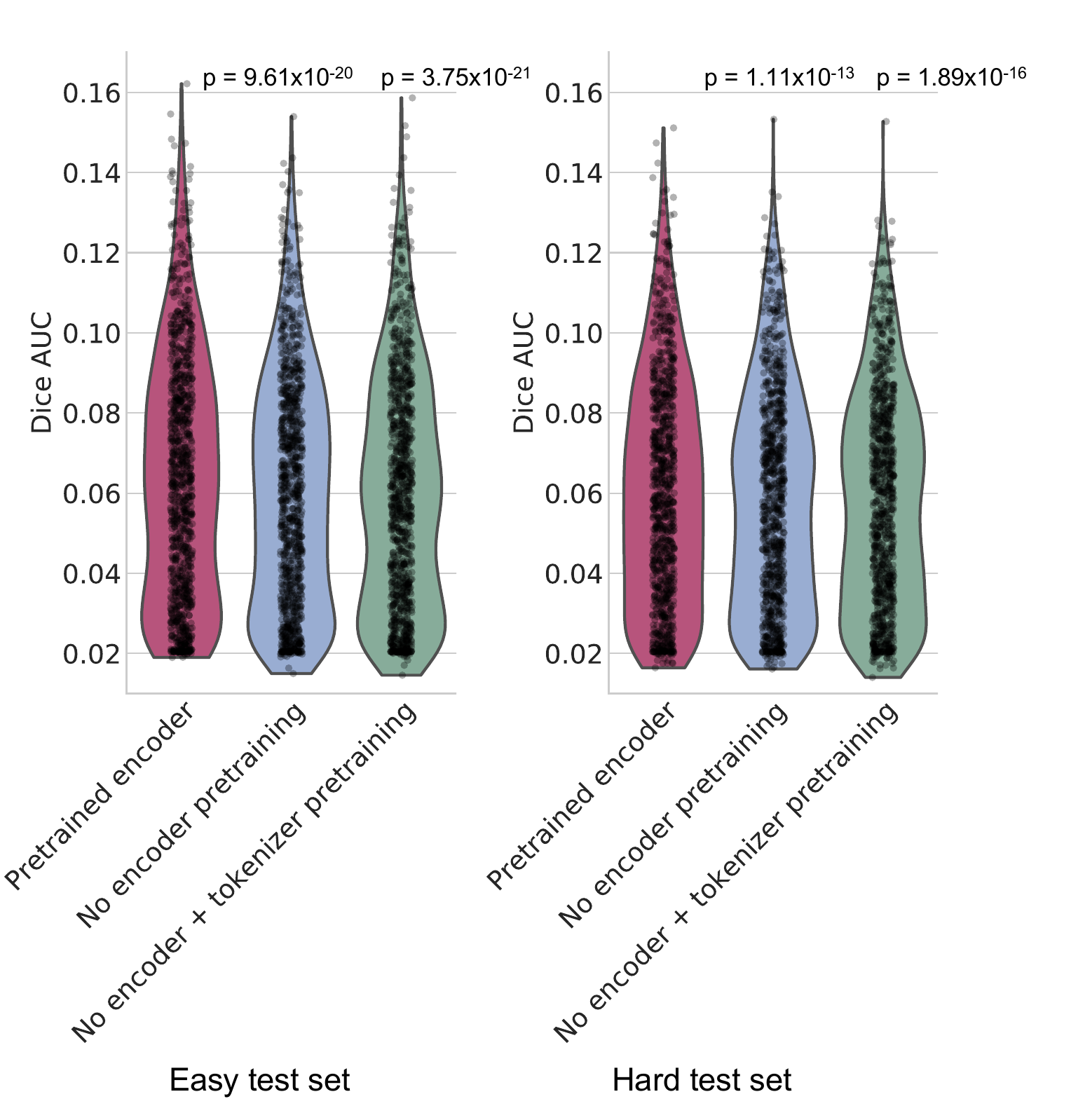}
 \caption{Effect of pretraining on predicting article-average activation maps from the articles' titles in the two test sets of Section 3.1.
 The text encoder of the full model was pretrained on non-neuroimaging articles.
 In the ``No encoder pretraining'' setup, the text encoder's weights were only trained on neuroimaging articles from the Neuroquery dataset, and not on non-neuroimaging articles.
 In the ``No encoder + tokenizer pretraining'' setup, the tokenizer was also not trained on non-neuroimaging articles.
 The p-values are estimated from paired-sample t-tests between the full setup with the text encoder's pretraining against the two setups without pretraining.}
 \label{fig:supp_ablation}
\end{figure}

\clearpage

\subsection{Effect of smoothness preprocessing}
\label{sec:ablation_fwhm}

\begin{table}[h]
\begin{tabular}{ | l | l | } 
\hline
 \makecell[l]{FWHM of smoothing Gaussian spheres}  & Mean AUC \\ \hline
 5 mm                                                    & 0.491   \\
 \makecell[l]{\textbf{9 mm}}                              & 0.507    \\
 15 mm                                                    & 0.474  \\
 \hline
\end{tabular}
\end{table}

\begin{figure}[ht!]
 \includegraphics[width=\linewidth]{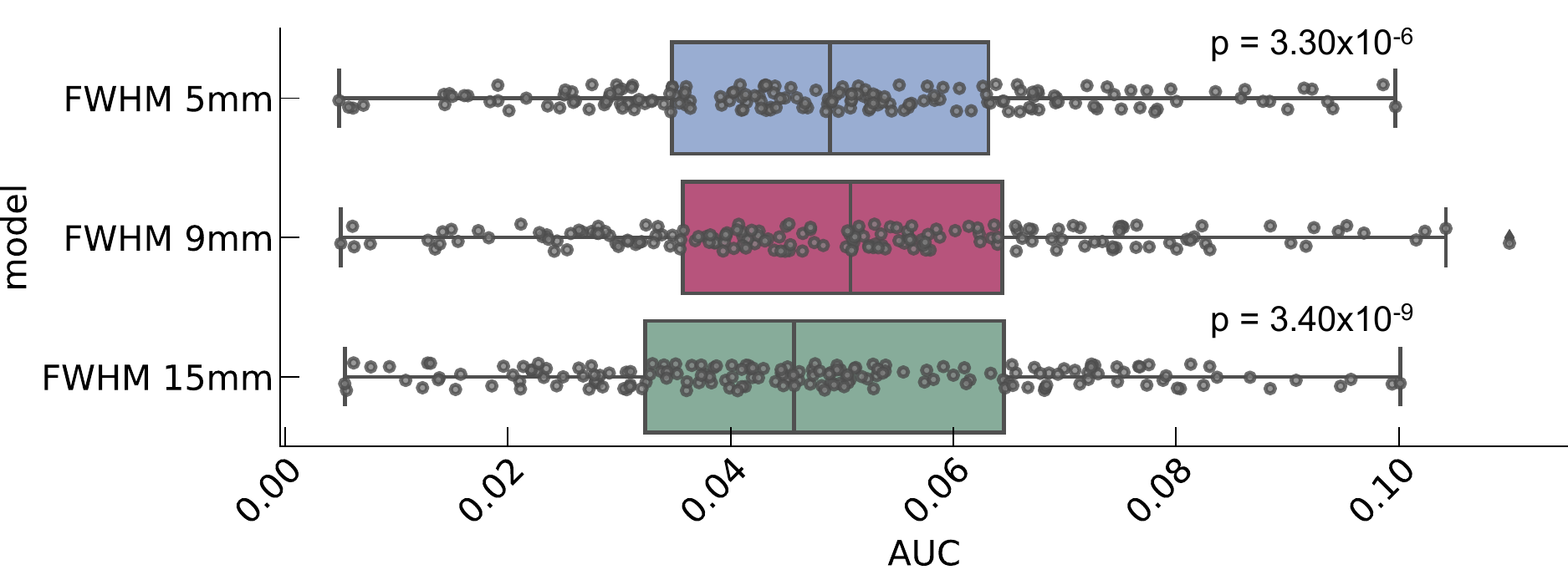}
 \caption{Dice AUCs of predicted IBC task activation maps from contrasts' description by Text2Brain with different full-width half-max (FWHM) of the Gaussian spheres used for preprocessing the Neuroquery training data. The p-values are estimated from paired-sample t-tests between the 9mm-FWHM setup against the two 5mm-FWHM and 15mm-FWHM setups.}
 \label{fig:supp_ablation}
\end{figure}

\subsection{Effect of HCP contrast's description length on model predictive accuracy}
\label{sec:ablation_fwhm}

\begin{table}[h]
\begin{tabular}{ | l | l | l | } 
\hline
 Model                     & Input type   & Mean AUC \\ \hline
 \multirow{2}{*}{Text2Brain} & IBC (short)  & 0.083    \\
                           & HCP (long)   & 0.076 ($\Delta = -0.007$)    \\ \hline
 \multirow{2}{*}{Neuroquery} & IBC (short)  & 0.078    \\
                           & HCP (long)   & 0.061 ($\Delta = -0.017$)    \\ \hline
 \multirow{2}{*}{Neurosynth} & IBC (short)  & 0.072    \\
                           & HCP (long)   & 0.067 ($\Delta = -0.005$)    \\ \hline
\end{tabular}
\caption{Dice AUCs of predicted HCP task activation maps from contrasts' IBC-based description and HCP-based description. $\Delta$s are the differences in Dice AUC between the two types of input.}
\end{table}

\begin{table*}[h]
\begin{tabular}{ |p{0.4\textwidth}|p{0.5\textwidth}| } 
\hline
 Task & Contrast \\ \hline
 \multirow{2}{0.4\textwidth}{Language Processing task consists of two runs that each interleave 4 blocks of a story task and 4 blocks of a math task. The goal of including the math blocks was to provide a comparison task that was attentionally demanding, similar in auditory and phonological input, and unlikely to generate activation of anterior temporal lobe regions involved in semantic processing, though likely to engage numerosity related processing in the parietal cortex.}
        & STORY: The story blocks present participants with brief auditory stories (5–9 sentences) adapted from Aesop’s fables, followed by a 2-alternative forced-choice question that asks participants about the topic of the story. \\
          & MATH: The math task presents trials auditorily and requires subjects to complete addition and subtraction problems. The trials present subjects with a series of arithmetic operations, followed by “equals” and then two choices. Participants push a button to select either the first or the second answer. The math task is adaptive to maintain a similar level of difficulty across participants. \\ \hline
\multirow{2}{0.4\textwidth}{RELATIONAL PROCESSING task localizes activation in anterior prefrontal cortex in individual subjects. The stimuli are 6 different shapes filled with 1 of 6 different textures.}
        &	REL: In the relational processing condition, participants are presented with 2 pairs of objects, with one pair at the top of the screen and the other pair at the bottom of the screen. They are told that they should first decide what dimension differs across the top pair of objects (shape or texture) and then they should decide whether the bottom pair of objects also differ along that same dimension (e.g., if the top pair differs in shape, does the bottom pair also differ in shape). \\
        &	MATCH: In the control matching condition, participants are shown two objects at the top of the screen and one object at the bottom of the screen, and a word in the middle of the screen (either “shape” or “texture”). They are told to decide whether the bottom object matches either of the top two objects on that dimension (e.g., if the word is “shape”, is the bottom object the same shape as either of the top two objects). \\ \hline
\multirow{2}{0.4\textwidth}{Social Cognition (Theory of Mind) is an engaging and validated video task was chosen as a measure of social cognition, given evidence that it generates robust task related activation in brain regions associated with social cognition and is reliable across subjects. } & Theory of Mind: an interaction that appears as if the shapes are taking into account each other’s feelings and thoughts. \vspace{1cm} \\
	& Random: there is no obvious interaction between the shapes and the movement appears random). \\ \hline
WORKING MEMORY task embeds the category specific representations component within the working memory task, by presenting blocks of trials that consisted of pictures of faces, places, tools and body parts.	& [stimulus type] contrast: [stimulus type] vs. fixation, collapsing across memory load

[stimulus type] vs AVG: stimulus type versus all other stimulus types

[stimulus type] can be one of the following ``body, faces, places, tools''  \\ \hline
MOTOR task identifies effector specific activations in individual subjects. Participants are presented with visual cues that ask them to tap their left or right fingers, squeeze their left or right toes, or move their tongue to map motor areas. & [movement type]: linear contrasts were computed to estimate activation for [movement type] versus baseline

[movement type – AVG]:  linear contrasts were computed to estimate activation for [movement type] versus all other movement types

[movement type] can be one of the following ``left hand, right hand, left foot, right foot, tongue''
 \\ \hline

\end{tabular}
\label{tab:long_description}
\caption{Long descriptions of Human Connectome Project (HCP) task contrasts. The contrast descriptions are constructed from the original full-text in~\citep{barch2013function}. We tried to stay consistent with the original text and only included contrast descriptions that do not require significant editing of the original text.}
\end{table*}

\begin{figure}[ht!]
 \includegraphics[width=\linewidth]{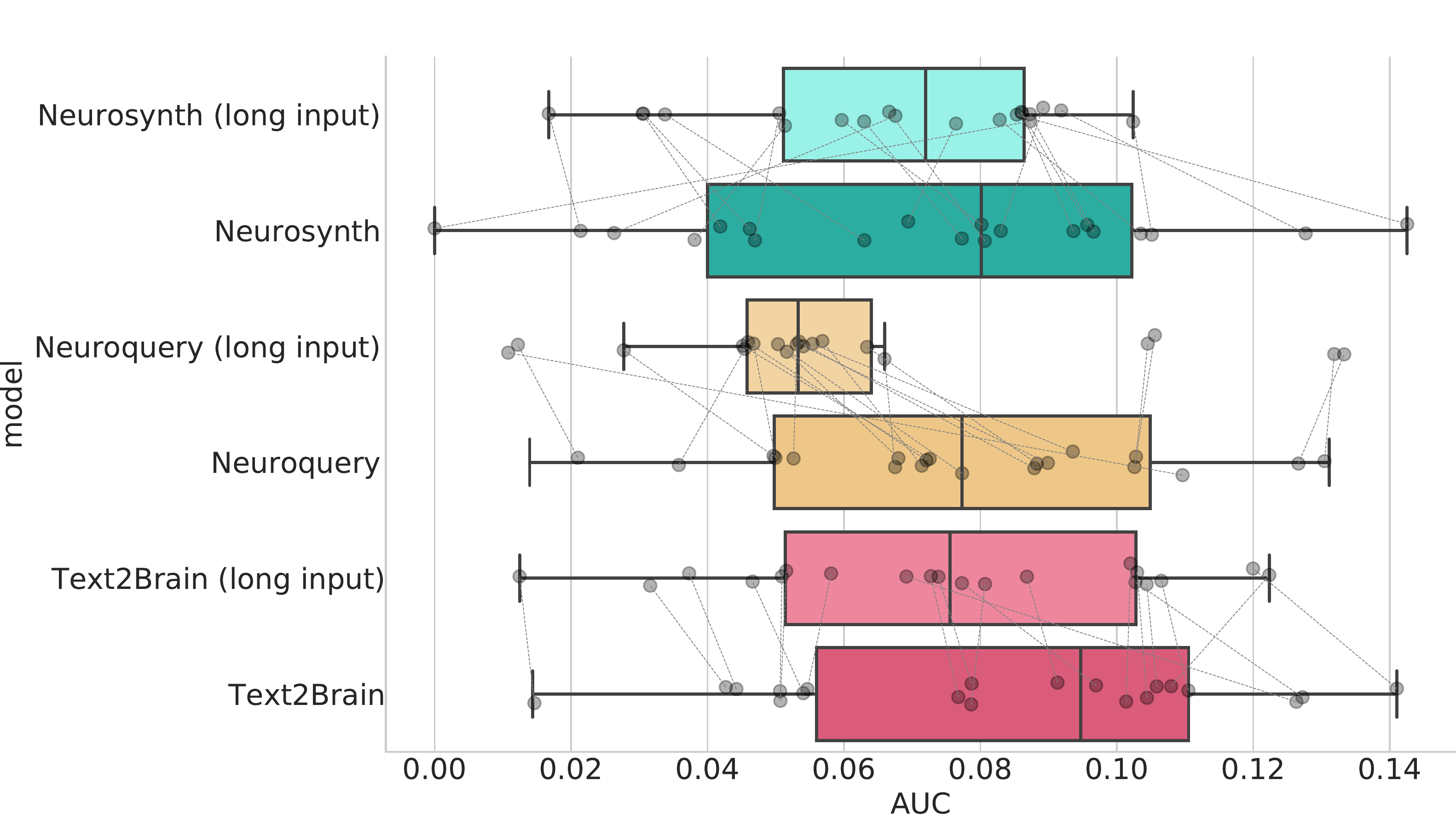}
 \caption{Comparison of predicted HCP task activation maps from contrasts' IBC-based description and HCP-based description (Table~\ref{tab:long_description}).}
 \label{fig:supp_long_input}
\end{figure}

\end{document}